\documentclass{kapproc} 

\usepackage{t1enc}

\usepackage{procps} 

\usepackage[dvips]{graphicx}

\upperandlowercase

\kluwerbib 

\begin{document}

\articletitle{A recent rebuilding of most spirals ?}


\author{Fran\c{c}ois Hammer\altaffilmark{1}, Hector Flores\altaffilmark{1}, Xianzhong Zheng\altaffilmark{2} and Yanchun Liang\altaffilmark{3}}

\altaffiltext{1}{Laboratoire Galaxies, Etoiles, Physique et Instrumentation\\
Observatoire de Paris, 92195 Meudon, France}
\altaffiltext{2}{Max-Planck Institut fuer Astronomie, Germany}
\altaffiltext{3}{National Astronomical Observatories, CAS, China}



\begin{abstract}
Re-examination of the properties of distant galaxies leads to the evidence that most present-day spirals have built up half of their stellar masses during the last 8 Gyr, mostly during several intense phases of star formation during which they took the appearance of luminous infrared galaxies (LIRGs). Distant galaxy morphologies encompass all of the expected stages of galaxy merging, central core formation and disk growth, while their cores are much bluer than those of present-day bulges. We have tested a spiral rebuilding scenario, for which 75$\pm$25\% of spirals have experienced their last major merger event less than 8 Gyr ago. It accounts for the simultaneous decreases, during that period, of the cosmic star formation density, of the merger rate, of the number densities of LIRGs and of compact galaxies, while the densities of ellipticals and large spirals are essentially unaffected.  
\end{abstract}


\section{Towards robust evolutionary features from z=1 to z=0} 
Here we summarize a study (Hammer et al, 2004, hereafter H04) which is based on a considerable amount of observations made using HST, ISO, VLA and VLT. It targets $\sim$ 200 galaxies (0.4$<$ z $<$1), mostly from the sample of the Canada France Redshift Survey (CFRS). CFRS is essentially complete up to z$\sim$ 1, encompassing all luminous ($M_{B}$ <- 20) galaxies with stellar mass ranging from 3 to 30 $10^{10}$ $M_{\odot}$ -hereafter called intermediate mass galaxies. Those account for 65\% to 80\% of the present-day stellar mass (Brinchman and Ellis, 2000; Heavens et al, 2004). Our goals were to collect significative evolutionary features since z=1, by combining:\\
1- Robust estimates of extinction, star and effective formation rates (SFRs and SFR/$M_{star}$) and O/H abundances at z$\sim$ 0.7. Those quantities have been derived after a detailed comparison of mid-IR and extinction corrected Balmer emission lines fluxes which provides consistent SFR estimates within a factor 2 (Flores et al, 2004; Liang et al, 2004a). Notice that Balmer emission lines have been properly corrected from the underlying absorption (based on high S/N spectra at moderate resolution, see Liang et al, 2004b). Notice also that OII or UV fluxes underestimates SFR by factors averaging to 5 to 22 for starbursts and LIRGs, respectively. \newline
2- A simplified morphological classification (see Zheng et al, 2004) which account for E/S0 and spirals as single classes,  while another class is assessed to objects barely resolved by the HST ($r_{half}<$ 3.5 $h_{70}^{-1}$kpc luminous compact galaxies, hereafter called LCGs). Our method, also based on color maps, considerably limits the uncertainties related to cosmological dimming, spatial resolution and morphological k-correction. It allows a fair comparison with morphological classification (derived at the same rest-frame wavelengths) of local galaxies in the same mass range (Nakamura et al, 2004).
\section{A formation history with violent IR episodes at z$<$ 1}
H04 found that $\sim$ 15\% of intermediate mass galaxies at z$>$ 0.4 are indeed luminous IR galaxies (LIRGs), a phenomenon far more common than in the local Universe. This is confirmed by preliminar Spitzer results (Yan et al, this volume). The high occurrence of LIRGs is easily understood only if they correspond to episodic peaks of star formation, during which galaxies are reddened through short IREs (infrared episodes). We estimate that each galaxy should experience 4 to 5 $\times$ $(\tau_{\rm IRE}/0.1Gyr)^{-1}$ IREs from z=1 to z=0.4. The star formation in LIRGs is sufficient in itself to produce 38\% of the total stellar mass of intermediate mass galaxies and then to account for most of the reported stellar mass formation since z=1. This is not surprising since integrations of the cosmic star
formation rate density, {\it if and only if they account for IR emission} (Flores et al., 1999;
 Elbaz and Cesarsky, 2003), match well the evolution of the global stellar mass density since z=1
(e.g. Dickinson et al. 2003; Heavens et al, 2004). It can be considered robust that 45$\pm$15\% of the mass
locked in present-day stars actually condensed in stars at $z <
1$. It is further supported by the luminosity-metallicity relation of z$\sim$ 0.7 emission line galaxies, which is found to be on average, metal deficient by a factor of $\sim$ 2 when compared to those of local spirals (Liang et al, 2004a, H04).\newline
A star formation history with short infrared episodes for most galaxies at intermediate redshifts is consistent with a hierarchical galaxy formation scenario. Observations of LIRG morphologies are then revealing us the physical processes which are responsible of most of the stellar mass production since z=1. Irregulars, major mergers and compact galaxies represent together about a third of the z$\sim$0.7 galaxy population and they almost disappear today (see Table 1 and also Lilly et al, 1998). Most of the star formation has occurred in them since they represent almost two-third of z$\sim$0.7 LIRGs (Table 1). We assume in the following that violent infra-red episodes are responsible of most morphological changes, which links in a simple way distant galaxies to those of the present-day Hubble sequence.   
  Which present day galaxy types are related to both morphological changes and episodic violent star formation events ? It cannot be present-day ellipticals because: (1) no LIRGs show an E/S0 morphology; (2) if the fate of all LCGs (or all LIRGs) were to become ellipticals, the density of present-day ellipticals would be much larger than observed (see Table 1). It leads H04 to assume that most of the recent star formation has occurred in progenitors of the numerous present-day spirals (see also Wolf et al, 2004). This is further supported by Zheng et al (2004a,b) who show that most z$\sim$ 0.7 spirals have blue cores (see also Ellis et al, 2001).

\begin{table}
\caption{From H04: morphological classification statistics for intermediate mass galaxies; z $\sim$ 0.7 galaxies are compared to those of the SDSS (Nakamura et al, 2004).}
\begin{tabular}{lccc} \hline
Type   & z$\sim$0.7 & z$\sim$0.7 & local \\ 
       & LIRGs &  galaxies & galaxies \\ \hline
E/S0        &   0\% & 23\% & 27\%\\
Spiral    &  36\% &  43\% & 70\%\\
LCG  &  25\% &  19\% & $<$ 2\%\\
Irregular  &  22\%  &  9\% & 3\%\\
Major merger &  17\% &  6\% & $<$ 2\%\\ \hline
\end{tabular}
\end{table}

\begin{figure}[ht]
\includegraphics[width=0.9\textwidth]{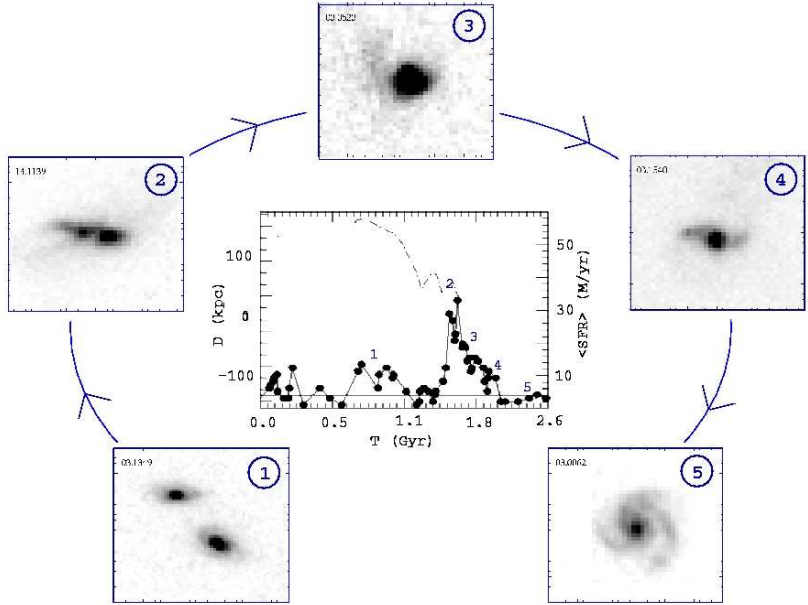}

\caption{A sequence of z$\sim$ 0.7 galaxies illustrating the rebuilding scenario; (1 \& 2): interaction and merging, (3 \& 4): post-merger prior to the disk re-building and (5): disk growth. (Middle): the star formation history derived by hydrodynamical simulations of a major merger (from Tissera et al, 2002) which shows the corresponding phases.75$\pm$25\% of present-day spirals have likely experienced this sequence since z=1.}
\end{figure}
\section{A scenario accounting for all evolutionary features ?}
The fraction of major mergers is evolving rapidly (Table 1; see also Conselice et al, 2004). Major mergers in our sample are showing two well identified nuclei separated by less than a galactic radius (Zheng et al, 2004a,b). Post mergers are naturally expected to show compact morphologies, and we assume in the following that LCGs are merger remnants. Indeed this was firstly claimed by Hammer et al (2001) on the basis of their detailed spectral properties. Moreover, Ostlin et al (2001) argued that their local counterparts are merger remnants on the basis of their velocity fields which are non rotationally supported. Because pair counts evolve very rapidly, the fraction of intermediate mass galaxies experiencing a major merger event since z=1 is estimated to be 0.75 on the basis of Bundy et al (2004), a value lower than what was found by Le F\`evre et al (2000). We then develop a scenario for which most spirals have been rebuilt during the last 8 Gyrs (Figure 1). H04 find that assuming the characteristic time for each phases (major mergers, compact and spirals) from hydrodynamical models (Tissera et al, 2004; see also Cox et al, 2004), one is able to reproduce the observed fraction of each species if 75$\pm$25\% of spirals were rebuilt since z=1. This simple scenario is consistent with 8 robust evolutionary features, namely the evolution of the global stellar mass, the L-Z diagram, the pair statistics, the IR light density, the spiral core colours, the number density of E/S0, Spirals, and peculiar galaxies (mergers and compact).\newline
A scenario showing no significant star formation in intermediate mass galaxies (Cowie et al, 1996; Brinchman \& Ellis, 2000) is inconsistent with 6 of these evolutionary features (stellar mass, L-Z, pair statistics, IR light density, spiral core colours and number density of peculiar galaxies). A scenario for which the stellar mass formation is dominated by minor encounters ("collisional starbursts", Somerville et al, 2001), is better, but still inconsistent with 4 evolutionary features (IR light density, number density of peculiar galaxies, spiral core colours and difficult to reconcile with pair statistics). In fact accounting for the IR light evolution leads to our spiral rebuilding scenario.\newline
This scenario has the advantage to be predictive and it can be tested. For example, the formation of a disk after a merger event requires that spiral progenitors have a substantially larger gas fraction than present-day spirals ($\sim$ 10-20\% of the baryonic mass). The considerable amount of stars formed during IREs suggests a gas content $\sim$ 5 times higher in z$\sim$ 0.7 galaxies. This is consistent with the reported O/H deficiency of z$\sim$0.7 galaxy gaseous phases (H04; Liang et al, 2004a). An important assumption is that LCGs are merger remnants. Preliminar results from Puech et al (2004, in preparation) and from Bershady et al (this volume) indicate that most LCGs have velocity fields not supported by rotation, as those studied locally by Ostlin et al (2001).\newline 
A spiral rebuilding scenario is very efficient in forming large bulges including those of the numerous early type spirals (75\% of present day spirals with intermediate masses). Robust conclusions on the past formation history can be derived on only 2 intermediate mass galaxies, e.g. Milky Way and M31. Recent studies of stellar populations in the M31 halo are suggestive of a complex formation history possibly with a recent merger (Brown et al, 2003; Rich, 2004). On the other hand, the Milky Way shows no trace of such a recent event. It is less bulge-dominated than M31, and we suggest that it could be part of the $\sim$ 25\% of galaxies which escaped a recent major merger at z $<$ 1.\newline
IR data reveal a strikingly different history than previous studies based on optical/UV data. Indeed the LIRG number density increases by factors up to 35 from z=0 to z=1 (Elbaz and Cesarsky, 2003), i.e. it evolves much rapidly than the UV luminosity density. H04 find that the optical properties of LIRGs mimic those of other galaxies, and that besides IR photometry, only detailed spectroscopic studies at moderate resolution are able to distinguish them.



\noindent



\noindent



\begin{chapthebibliography}{1}
\bibitem{} Brinchmann, J. \& Ellis, R. S.  ApJ  536, L77-L80 (BE2000, 2000)
\bibitem{} Brown, T. M., Ferguson, H. C., Smith, E., Kimble, R. A. et al. ApJ, 592, L17 (2003)
\bibitem{} Bundy, K., Fukugita, M., Ellis, R. S., Kodama, T. \& Conselice, C. J.  ApJ 601, 123 (2004)
\bibitem{} Cowie, L. L.; Songaila, A.; Hu, E., Cohen, J. Astron. J 112, 839 (1996)
\bibitem{} Dickinson, M. Papovich, C. Ferguson, H.C. Budavari, T.,  ApJ 587, 25D (2003)
\bibitem{} Elbaz, D. \& Cesarsky, C. J.  Science 300, 270-274 (2003)
\bibitem{} Ellis, R., Abraham, R. \& Dickinson, M.  ApJ 551, 111-130 (2001)
\bibitem{} Flores, H. Hammer, F., Thuan T.X. et al.  ApJ 517, 148-167 (1999)
\bibitem{} Flores, H., Hammer, F., Elbaz, D. et al. A\&A 415, 885-888 (2004)
\bibitem{} Hammer, F., Gruel, N.,Thuan, T. X., Flores, H. \& Infante, L.   ApJ 550, 570-584 (2001)
\bibitem{} Hammer, F., Flores, H., Elbaz, D., Zheng X. Z., Liang, Y. C.\& Cesarsky, C., A\&A, in press (H04, astro-ph/0410518)
\bibitem{} Heavens, A. Panter, B. Jimenez, R. Dunlop, J.,  Nature 428, 625 (H2004, 2004)
\bibitem{} Le F\`evre, O. et al. MNRAS 311, 565-575 (2000)
\bibitem{} Liang, Y. C., Hammer, F., Flores, H., Elbaz, D. \& Cesarsky, C. J.  A\&A 423, 867 (2004a)
\bibitem{} Liang, Y. C.,  Hammer, F.,  Flores, H., Gruel, N. \& Assemat, F.    A\&A 417, 905 (2004b)
\bibitem{} Lilly, S. J. et al. ApJ 500, 75-94 (1998)
\bibitem{} Nakamura, O. Fukugita, M. Brinkmann, J et al. AJ 127, 2511 (2004)
\bibitem{} Ostlin, G., Amram, P., Bergvall, N.et al.  A\&A 374, 800-823 (2001)
\bibitem{} Somerville, R. S., Primack, J. R. \& Faber, S. M., MNRAS 320, 504 (2001)
\bibitem{} Tissera, P. B.; Dominguez-Tenreiro, R.; Scannapieco, C.; Saiz, A.  MNRAS 333, 327 (2002)
\bibitem{} Wolf, C., Bell, E., McIntosh, D., Rix,H.W. et al., ApJ submitted (astro-ph/0408289)
\bibitem{} Zheng, X. Z., Hammer, F., Flores, H., Assemat, F. \&  Pelat, D.  A\&A 421, 847 (2004)
\bibitem{} Zheng, X. Z., Hammer, F., Flores, H., Assemat, F., Rawat, A., A\&A (submitted) (2004b)

\end{chapthebibliography}
\end{document}